# Raman 2D-Band Splitting in Graphene: Theory and Experiment

*Otakar Frank[1,2], Marcel Mohr[3], Janina Maultzsch[3], Christian Thomsen[3], Ibtsam Riaz[4], Rashid Jalil[4], Kostya S. Novoselov[4], Georgia Tsoukleri[1], John Parthenios[1], Konstantinos Papagelis[1,5*], Ladislav Kavan[2] and Costas Galiotis[1,5]*

[1]Institute of Chemical Engineering and High Temperature Chemical Processes, Foundation of Research and Technology-Hellas (FORTH/ICE-HT), 265 04 Patras, Greece

[2]J. Heyrovsky Institute of Physical Chemistry, v.v.i., Academy of Sciences of the Czech Republic, 182 23 Prague 8, Czech Republic

[3]Institut für Festkörperphysik, Technische Universität Berlin, 106 23 Berlin, Germany

[4]School of Physics and Astronomy, University of Manchester, Manchester M13 9PL, UK

[5]Materials Science Department, University of Patras, 265 04 Patras, Greece

* Corresponding author: kpapag@upatras.gr

ABSTRACT

We present a systematic experimental and theoretical study of the two-phonon (2D) Raman scattering in graphene under uniaxial tension. The external perturbation unveils that the 2D mode excited with 785nm has a complex line-shape mainly due to the contribution of two distinct double resonance scattering processes (inner and outer) in the Raman signal. The splitting depends on the direction of the applied strain and the polarization of the incident light. The results give new insight into the nature of the 2D band and have significant implications for the use of graphene as reinforcement in composites since the 2D mode is crucial to assess how effectively graphene uptakes an applied stress or strain.

KEYWORDS

Graphene, Raman spectroscopy, tensile strain, 2D mode

Graphene is the thinnest known elastic material, exhibiting exceptional mechanical and electronic properties.[1] Because graphene is a single-layer membrane, it is also amenable to external perturbations, including mechanical loading. A promising approach to develop graphene-based electronic devices is by engineering local strain profiles obtained by means of a controlled mechanical or thermal deformation of the substrate or by applying appropriate geometrical patterns in a substrate. The strained material becomes a topological insulator enabling the opening of significant energy gaps in graphene's electronic structure.[2, 3] An alternative approach in this line is to control and manipulate the intrinsic ripples in graphene sheets using thermally generated strains.[4]

This feature is expected to strongly influence electronic properties by inducing effective magnetic fields and modify local potentials.[5] Thus, the precise determination and monitoring of strain is an essential prerequisite for understanding and tuning the interplay between the geometrical structure of graphene and its electronic properties.

Raman spectroscopy is a key diagnostic tool to identify the number of layers in a sample and probe physical properties and phenomena.[6-9] The G band is the only Raman mode in graphene originating from a conventional first order Raman scattering process and corresponds to the in-plane, zone center, doubly degenerate phonon mode (transverse (TO) and longitudinal (LO) optical) with $E_{2g}$ symmetry[10]. The D and 2D modes come from a second-order double resonant process between non equivalent K points in the Brillouin zone (BZ) of graphene, involving two zone – boundary phonons (TO-derived) for the 2D and one phonon and a defect for the D band.[11] Both modes are dispersive spectral features, *i.e.* their frequencies vary linearly as a function of the energy of the incident laser, $E_{laser}$. The slope $\partial \omega_{2D} / \partial E_{laser}$ is around 100cm$^{-1}$/eV, which is approximately twice of that of the D peak.

Recent experiments [12-14] assessed the strain sensitivity of the G and 2D bands under uniaxial tensile and compressive strains and revealed that reversible and controlled deformation can be imparted to graphene. In general, tensile strain induces phonon softening (red shift) and compression causes phonon hardening (blueshift). In addition, the G peak splits into two components due to symmetry lowering of the crystal lattice.[12, 13, 15] The sub-peaks denoted G$^-$ and G$^+$ (in analogy to carbon nanotubes) shift at rates ~30 and ~10 cm$^{-1}$/%, respectively, both under uniaxial tension[12, 13] and compression.[12] Mohiuddin *et al.*[13] and Tsoukleri *et al.*[14] observed a large variation of the graphene 2D band with strain at a rate higher than ~50 cm$^{-1}$/% using the 514.5 nm (2.41 eV) excitation. Comparable results were obtained using the 785 nm (1.58 eV) excitation.[12]

Also, biaxial strain induces shift of the Raman peaks at rates approximately double than those observed in uniaxial tension.[16-18] Several other studies reported significantly lower shift rates for both G and 2D peaks under uniaxial tension.[19-22] However, given the results above, which are in excellent agreement with recent theoretical calculations,[13, 15] the $G^-$ and $G^+$ shift rates of ~30 and ~10 cm$^{-1}$/% should be used as the benchmark for further strain assessment. On the other hand, the 2D peak splitting of graphene having zigzag and armchair orientations with respect to the strain axis has been quite recently reported in Ref. 19 using the 532 nm (2.33 eV) excitation. Samples with intermediate orientation showed strain induced broadening but no splitting.[19] The phenomenological interpretation[19] of the processes responsible for the observed splitting is different from the results presented in this article (*vide infra*).

In addition to the above, Raman spectroscopy has been extensively used to assess the degree of stress/strain transfer and the role of the interface in a variety of composites reinforced with carbon–based materials such as carbon fibres,[23, 24] graphene[25] and various types of carbon nanotubes.[26] The carbon reinforcements exhibit relatively strong Raman signal, enabling the stress-transfer to be monitored through optically transparent matrices. Although the 2D peak exhibits a much broader linewidth compared to the G peak, its high Raman shift rate allows even small stress/strain variations to be easily detected, making this mode highly sensitive upon application of tensile or compressive stress/strain. Therefore, the response of the 2D band under various excitation wavelengths will have important implications for the use of graphene as reinforcement in polymer composites.

In this work, we have investigated the 2D band evolution in single layer graphene flakes, with three different crystallographic orientations, for tensile strains up to ~1% at two different incident laser polarizations (parallel and perpendicular to the strain axis). The experiments have been conducted with the 785 nm excitation. In contrast to previous results obtained with 514.5 nm excitation, the 2D feature shows a remarkable response upon strain, namely an extensive asymmetric peak broadening, eventually leading to a clear splitting at higher strain levels. As will be discussed further on, the 2D mode splitting cannot be explained only by strain induced asymmetry of the graphene's BZ, instead both the inner and outer scattering mechanisms have to be taken into account.[27] From first principles calculations we investigate the scattering along the lines connecting neighboring K-valleys, distinguishing into inner and outer processes (see below). This leads to at least six channels which contribute to the 2D-mode, and whose frequencies are in general different. The calculated individual shift values have been used to simulate the Raman spectra of the studied flakes for two different polarization directions of the incident light. The qualitative agreement between the experimental and simulated spectra gives solid evidence for the proposed origin of the 2D mode.

RESULTS AND DISCUSSION

Graphene monolayers were subjected to tensile loading by means of a cantilever beam assembly.[12, 14] The specimens were embedded into two polymeric layers of SU8 and S1805 photoresists and placed onto PMMA bars (see Methods and Refs. 12, 14). Raman measurements were performed *in situ* on different sample locations, which were chosen after a detailed Raman mapping of the flake area (crosses in Figure S1 in Supporting Information).

Figure 1 shows representative Raman spectra of a graphene monolayer (flake F1) in the 2D mode region as a function of strain and laser polarization (parallel and perpendicular to the strain axis). Upon tensile strain the 2D feature shifts towards lower frequencies, accompanied with a significant asymmetric broadening, which becomes more pronounced at higher strain levels. From Fig. 1 it is evident that the 2D peak can be deconvoluted into two components even at zero strain conditions, for both laser polarizations. A superposition of two Lorentzian components with unconstrained widths fits excellently the Raman spectra. The detected splitting of the 2D mode increases with the strain level. At higher strain levels, the 2D mode can be fitted by more than two Lorentzians, although the fit quality is improved only marginally. It should be stressed that the G mode of the flake F1 excited at 785 nm splits in two components polarized along ($G^-$) and perpendicular ($G^+$) to the strain axis.[12] The strain sensitivities of the individual G peaks under tension is -31.4 ± 2.8 $cm^{-1}$/% for the $G^-$ mode and -9.6 ± 1.4 $cm^{-1}$/% for the $G^+$. These values are in excellent agreement with previous tensile measurements, at 514.5 nm excitation, on bare graphene performed at extremely small strain rates.[13] It should be noted that the parallel (Figure 1, left panel) and perpendicular (Figure 1, right panel) laser polarization experiments were not conducted in the same experimental run, therefore the initial 2D band position for the latter one is upshifted by ~3 $cm^{-1}$, suggesting the presence of small residual compressive strain in the monolayer. However, the G sub-bands evolution for the $\theta_{in} = 90°$ experiment is linear and with the same shift rates (within the experimental error) as in the $\theta_{in} = 0°$ case, hence we can assume the 2D band behavior for $\theta_{in} = 90°$ is representative as well. Crystallographic orientation w.r.t. the strain axis ($\varphi$) of the studied graphene flakes was determined using the intensities of $G^-$ and $G^+$ sub-peaks (for details of the procedure see Supporting Information).[12, 13, 20] When the angle $\varphi$ is defined from the zigzag direction (see Figure S2 in Supporting Information), the resulting orientations of the examined flakes are $\varphi_{F1} = 18.4 ± 0.1°$, $\varphi_{F4} = 10,8 ± 0.8°$ and $\varphi_{F5} = 23.2 ± 1.0°$.

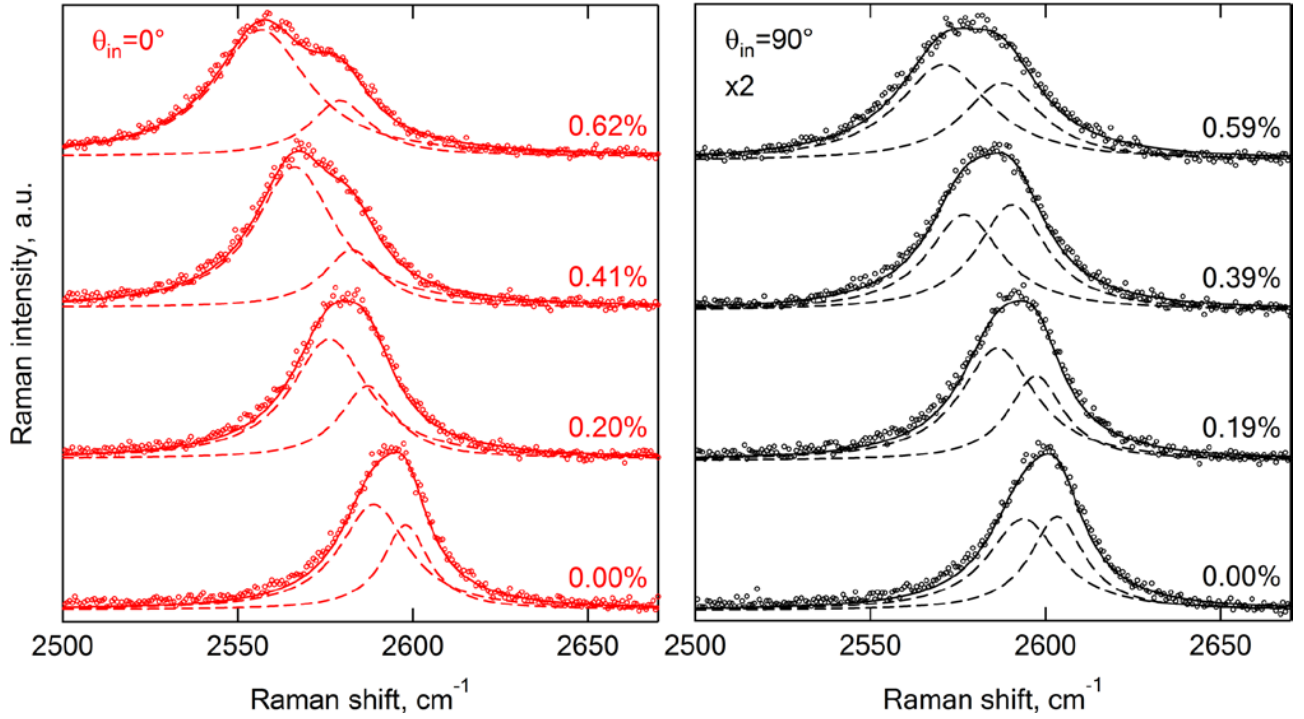

**Figure 1.** 2D mode Raman spectra of graphene flake excited at 785 nm under uniaxial tension. The spectra in the left and right panels were acquired with laser polarization parallel ($\theta_{in}=0°$) and perpendicular ($\theta_{in}=90°$) respectively, w.r.t. the strain axis. Data were recorded around the center of the flake F1. The original measurements are plotted as points. The solid curves are the best Lorentzian fits to the experimental spectra while the dashed lines indicate the two components (see text). The tension level is labeled at each curve.

It is well established that the 2D band in graphene is activated from a double resonance (DR) Raman scattering mechanism.[11, 28, 29] This band consists of an intervalley DR process which involves an electron or a hole in the vicinity of the Dirac K point and two in-plane transverse optical (TO) phonons around the K-point having $\boldsymbol{q_1}+\boldsymbol{(-q_1)}=\boldsymbol{0}$ (Figure 2). The contributing TO-derived phonon branch is the one with the highest energy along the K − M direction in the BZ direction. Although the 2D mode includes contributions from the entire Brillouin zone, we will focus here on those directions that connect one K-point with its three neighboring K-points (see Fig. 2).[27] Recently, we have undertaken a systematic theoretical study of the strain-induced modifications of the electronic bands and phonon dispersions of graphene.[15] The application of uniaxial strain distorts the reciprocal lattice, as well as the positions of the high symmetry points in the BZ. For example, for strain along the $\varphi = 20°$ crystallographic direction (see Figure 2) the graphene point group symmetry is reduced from $D_{6h}$ to $C_{2h}$. The shaded areas in Fig.2(a) enclose the irreducible parts of the BZ. Therefore, the band structure of strained graphene ($\varphi = 20°$) should be plotted along BZ paths with unequal length such as $\Gamma - K_1 - M_1 - \Gamma$, $\Gamma - K_2 - M_2 - \Gamma$ and $\Gamma - K_3 - M_3 - \Gamma$; the electronic bands are changing differently along the different $K - K'$ paths ($K_3 - K_2$, $K_4 - K_3$, $K_1 - K_2$ in Fig.2(a)), with respect to the strain direction. Depending on which direction in reciprocal lattice the electron is scattered, different resonant conditions should apply and, consequently, different phonon wavevectors $q$ become enhanced by DR. Thus, the recorded 2D Raman band shift in uniaxially strained graphene is the combination of the 2D phonon softening due to strain and an additional shift due to the relative movement of the Dirac cones modifying the wave vector of the participating phonon.[13, 15, 27] The latter leads to differences in phonon frequency up to 10 cm$^{-1}$ per 1% strain[15] and seems to be the main source of 2D band broadening under strain, if only one of the so-called inner or outer processes contributes (*vide infra*). Finally, it should be stressed that our first principles calculations shows that no sizable energy band gap opens in uniaxially strained for small strain levels, in agreement with very recent calculations.[30]

Very recently, it has been shown by us[27] that both inner and outer processes (Fig. 2(b)) contribute to the Raman cross section of the 2D peak, even at zero strain. In the outer (inner) process the participating phonons satisfy DR conditions coming from $K - M$ ($\Gamma - K$) with momentum $q_{out} > q_{KK'}$ ($q_{in} < q_{KK'}$). Here, we attempt to provide experimental verification of the above predictions by comparing our experiments with the calculated findings for strain along the $\varphi = 10°$ and $20°$ directions which coincide closely with the orientations of the studied flakes. For this purpose, a detailed "mapping" of the 2D band frequencies, considering both the inner and outer processes, has been undertaken by conducting first principles calculations of the phonon dispersion and the electronic band structure for small uniaxial strains up to ~2%. The scattering of electrons is considered elastic in order to identify the involved phonon wavevector (this corresponds to the almost horizontal arrows in Fig. 2(b)). Figures 2(c) and S3 (Supporting Information) present the obtained 2D frequencies for the inner and the outer processes for strains along $20°$ and $10°$ directions, respectively, using the 785 nm laser excitation. According to the theoretical results, inner and outer processes give rise to two distinct frequencies separated by 9.2 cm$^{-1}$ for unstrained graphene (Fig. 2(c)). As presented in Fig.1 and S4 (Supporting Information), the 2D band shape at zero strain level ($\lambda_{exc}$ = 785nm) is clearly asymmetric for both polarization geometries and can be fitted adequately by using two Lorentzian components. The experimentally obtained frequency positions at zero strain differ, in all cases, by almost 10 cm$^{-1}$ in excellent agreement with the theoretical predictions. The upper (lower) frequency stems from the phonon branch between $\Gamma$ and K (K and M). As the strain increases, this results in three different K − K´ paths due to the induced asymmetry of the BZ for $\varphi = 20°$. These three paths give rise to three outer (o) and three inner (i) processes such as o($K_3 − K_2$), i($K_3 − K_2$), o($K_4 − K_3$), i($K_4 − K_3$), o($K_1 − K_2$), i($K_1 − K_2$), (see Fig. 2(a)) and as the equivalence between the three directions for this strain direction is lifted, six components with distinct frequencies emerge. Figures 2(c) and S3 (Supporting Information) show that each component follows a different strain dependence, while the inner process leads to a more pronounced splitting than the outer process. Within the framework of the present calculations it becomes apparent that the two distinct Lorentzian components, used to fit the 2D peak in Fig.1 (and S4, S6 in Supporting Information), is only a rough approximation of the actual 2D band splitting and broadening. However, as shown below, the decomposition of the 2D feature in two sub-peaks provide an adequate fit even to simulated spectra, where six components were used to compose the 2D peak. Finally, it should be stressed that the physical origin of the larger splitting for the inner processes compared to the outer ones can be traced to the strain-direction dependent softening of the TO-mode. The phonon branches corresponding to inner processes show a stronger strain-direction-dependence than the phonon branches corresponding to outer processes.[27]

The strain dependence of representative 2D Raman spectra for the flakes F4 and F5, using the 785nm excitation, is illustrated in Figs. S4 and S6, resp. (Supporting Information). Upon comparison with Fig. 1, it is clear that, for all flakes, the 2D spectral feature excited with the same laser polarization is quite similar in spite of the different crystallographic orientations of the samples. The small deviations, especially in the perpendicular polarization, are probably caused by the different lattice orientations. However, the divergence is not as pronounced as between the armchair and zig-zag configurations, which are the extreme cases.[19, 27]

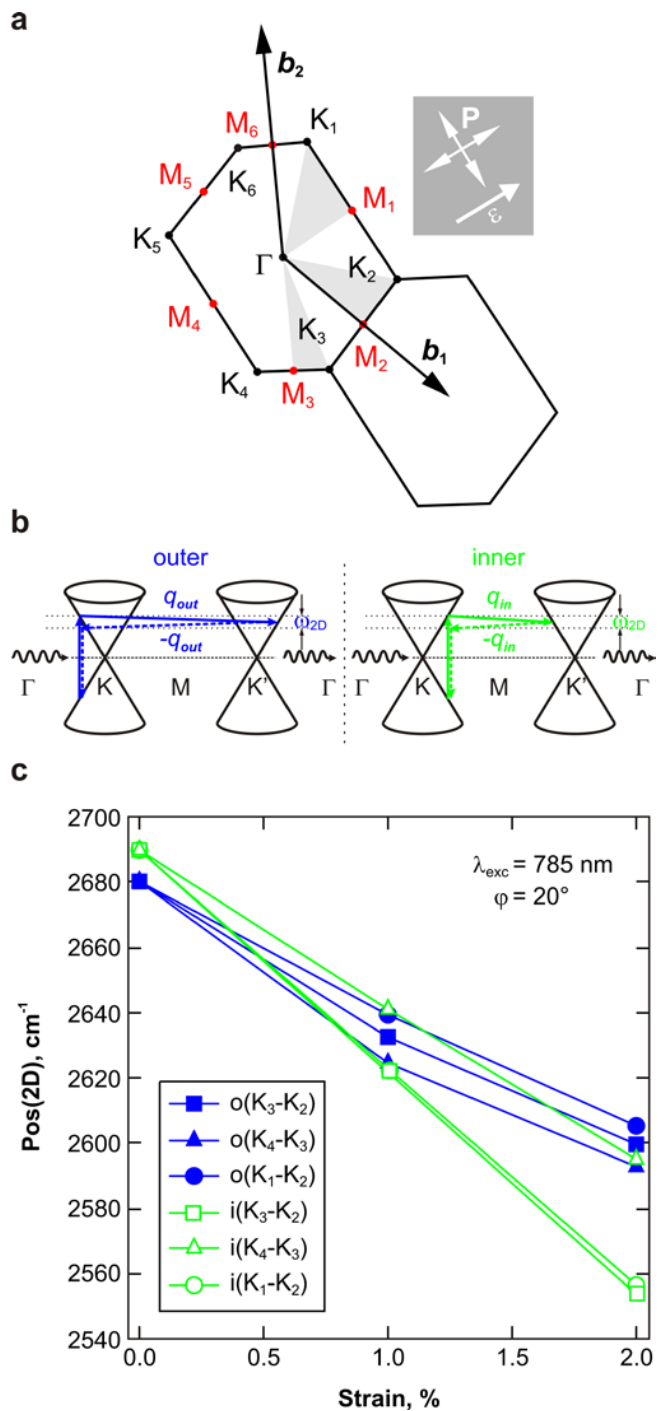

**Figure 2.** (a) The first Brillouin zone of uniaxially strained graphene for $\{\epsilon=0.3, \varphi = 20°\}$. The $\boldsymbol{b_1}$ and $\boldsymbol{b_2}$ denote the primitive vectors of the reciprocal lattice. The high symmetry points and the irreducible parts (shaded areas) of the Brillouin zone are depicted. (b) Incoming double resonance

mechanisms (left: outer process and right: inner process) plotted into the band structure scheme of graphene along the high symmetry KMK' path. For simplicity we omit the equivalent process for hole-phonon scattering. (c) Calculated 2D frequencies for the outer (o) and inner (i) processes along different K – K' paths (Fig. 2(a)), for laser excitation 785nm and $\varphi = 20°$. The blue solid (green open) symbols correspond to the outer (inner) process. The lines are guides to the eye.

Grüneis et al.[31] used first order time dependent perturbation theory and showed that the absorption probability per unit time $W(\mathbf{k})$ for an electron with wavevector $\mathbf{k}$ is proportional to $|\mathbf{P}\mathbf{k}|^2$, where $\mathbf{P}$ is the polarization vector of the incident light. This means that the polarization of the incoming laser affects the absorption in $\mathbf{k}$-space; it has a maximum for electrons with wave vector $\mathbf{k}$ perpendicular to $\mathbf{P}$, and it is null for electrons with $\mathbf{k}$ parallel to $\mathbf{P}$. Therefore, taking into account the incident laser polarization, electrons from specific areas in the reciprocal space cannot be excited and do not contribute to the Raman cross section for the 2D peak.[27] Mohr et al.[27] showed that both inner and outer processes lead to polarization dependent frequency shifts for the 2D mode.

As shown in Fig.1 (and Figs. S4 and S6, Supporting Information), the 2D band evolution with strain is highly sensitive to the incident light polarization. In fact, the processes perpendicular to the polarization are the most dominant ones, whereas the electrons with a wavevector parallel to the polarization of a photon cannot be excited during an absorption process. Here in order to estimate the effect of incident light polarization w.r.t. the strain axis on the 2D mode and the concomitant splitting under strain, we have performed a simulation of the Raman spectra for specific polarization conditions. Since the 2D peak is composed from overlapping sub-peaks we simulated the 2D spectra as a function of strain and laser polarization taking into account the variations in the shift rates of the individual processes (Fig.2(c) and Fig.S3 in Supporting Information). We assumed that the contributions of the processes originating from the three unequal K – K' paths are determined by the angle between the laser polarization direction and the particular K-K' line as sine squared.

Figure 3 shows the simulated Raman 2D mode spectra as a function of strain for the graphene layer with lattice orientation $\varphi = 20°$, which is a value very close to the orientation of the flake F1. The band evolution shows almost identical aspects as in the case of experimentally obtained spectra (Fig. 1). The simulation was performed using the calculated frequencies of the six different processes for both inner and outer paths (o($K_3 - K_2$), i($K_3 - K_2$), o($K_4 - K_3$), i($K_4 - K_3$), o($K_1 - K_2$), i($K_1 - K_2$)) for 785 nm excitation. The amplitude for each of the six sub-peaks was determined by the angle between the laser polarization direction and the particular K-K' line as sine squared. Thus, the maximum amplitude of 1 is set for a process perpendicular to the laser polarization. Based on a deconvolution of the measured 2D band at zero strain, the outer to inner processes ratio was estimated as approx. 1:2 for the simulations and the FWHM of all the individual components as 20 cm$^{-1}$. Such initial peak width was found to provide the FWHM of the simulated 2D band of 24 cm$^{-1}$. However, changing the peak width alters neither the slopes nor the general shape of the constructed 2D peak. As can be seen from Fig. 3, the resulting simulated 2D band structure can be readily fitted by only two Lorentzian components as in the case of the experimental data in spite of being in fact convoluted from six Lorentzians (Fig. S8, Supporting Information). When comparing the experimental and simulated spectra (Figures 1 and 3, respectively), the peak shapes for a particular polarization evolve in the same fashion and in both cases the peak shapes, widths and strain shift rates essentially depend on the incident laser polarization. To further prove the agreement between experiment and theory, Figure S5 (Supporting Information) shows the simulated spectra calculated for a graphene flake with lattice orientation $\varphi = 10°$, which closely resembles the spectra measured on the flake F4 with $\varphi = 10.8°$ (Fig. S4, Supporting Information).

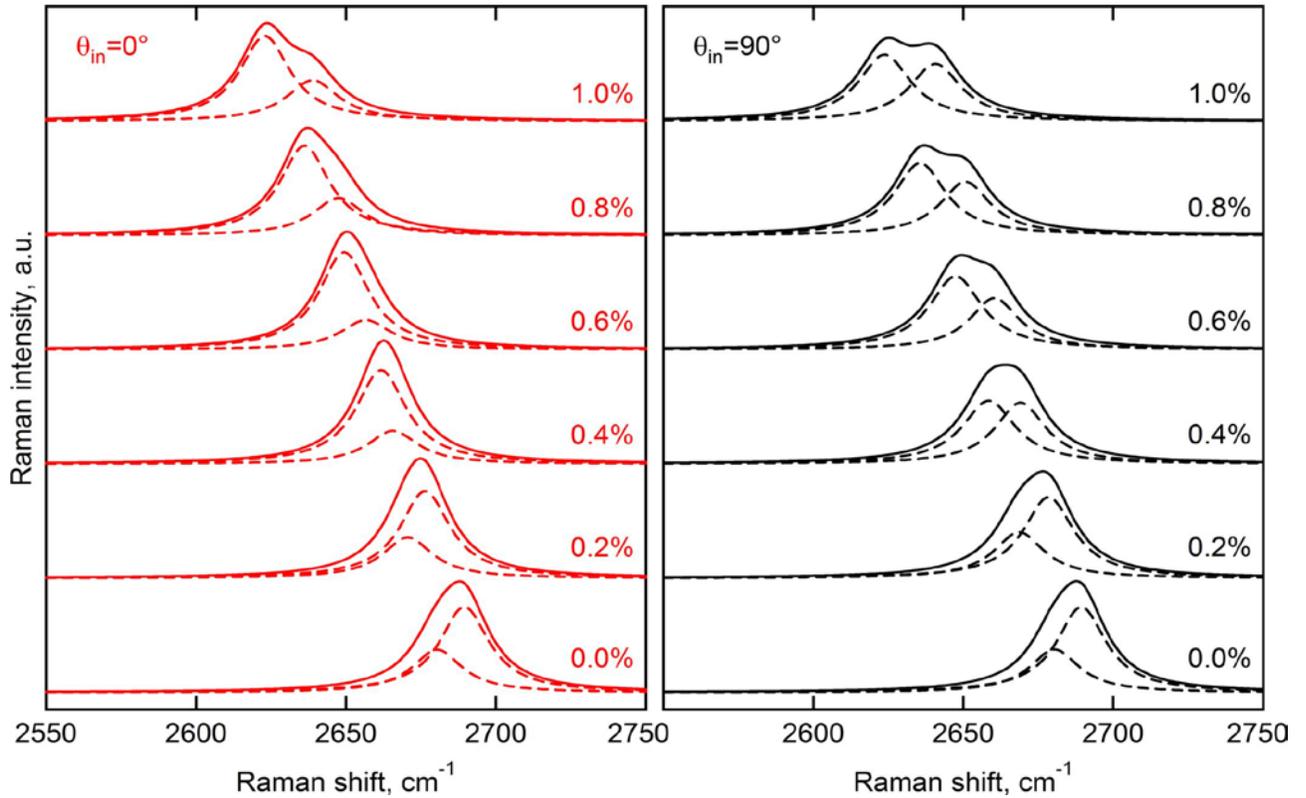

**Figure 3.** Simulated Raman spectra in the 2D frequency region calculated for 785 nm excitation of a graphene flake with crystal lattice orientation $\varphi = 20°$ and for various tensile strain levels. The spectra in the left and right panels were calculated for laser polarization parallel ($\theta_{in}=0°$) and perpendicular ($\theta_{in}=90°$) w.r.t. the strain axis, respectively. Dashed lines indicate fits with two Lorentzian components to the calculated spectra.

Figure 4 shows the experimentally extracted frequencies of the two main 2D Lorentzian components as a function of tensile strain for the flake F1. The lower frequency component depends linearly on strain (with a slope of -46.8 ± 2.1 cm$^{-1}$/%), whereas the higher frequency one exhibits a clearly non-linear trend accompanied by a slope change at a strain of about 0.2%. The initial slope of the latter sub-peak is quite similar to that exhibited by the lower frequency counterpart, while after the inflection point (~0.2%) the shift rate is reduced to -23.6 ± 3.6 cm$^{-1}$/%. The corresponding slopes of the simulated spectra for $\theta_{in}$=0° ($\theta_{in}$=90°) are -57.4 ± 1.5 cm$^{-1}$/% (-55.9 ± 0.5 cm$^{-1}$/%) for the lower component and -49.4 ± 1.3 cm$^{-1}$/% (-47.6 ± 0.5 cm$^{-1}$/%) for the higher one. The agreement between theory and experiment is excellent for the lower component whereas for the higher component there is good agreement only with the initial experimental slope (-44.7 cm$^{-1}$/%). Exactly the same behavior was observed in all tension experiments on a different flake F5 (Figs. S6 and S7, Supporting Information). The observed slope change at ~0.2 % is puzzling at this stage of research and detailed theoretical investigations are required to shed light on this issue.

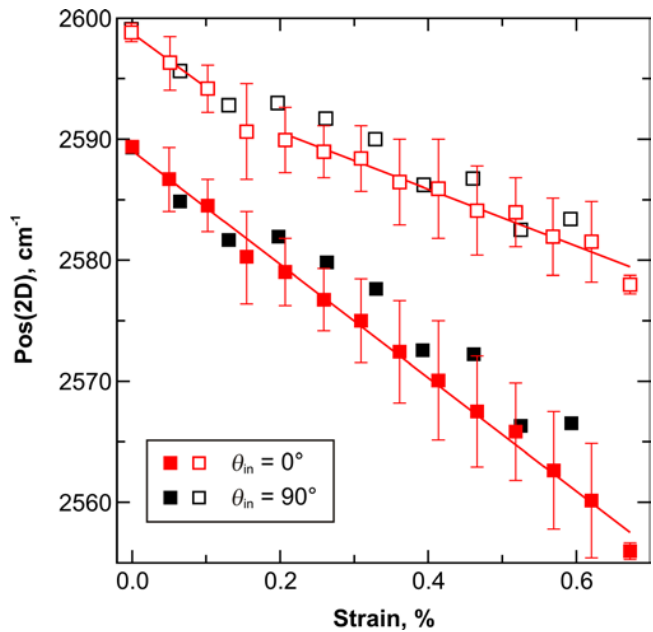

**Figure 4.** Splitting of the Raman 2D mode under tensile strain as measured on the graphene flake F1. Red and black colors indicate laser polarization parallel ($\theta_{in}$=0°) and perpendicular ($\theta_{in}$=90°)

w.r.t. the strain axis, respectively. Full and empty squares of the same color indicate the frequency of the two main Lorentzian components of the 2D mode. Solid lines (only for $\theta_{in}=0°$) represent linear fits where all measurements on the flake have been taken into account. The data points for $\theta_{in}=90°$ were offset by -4 cm$^{-1}$ in order to provide a better visual comparison to data acquired with $\theta_{in}=0°$. The original spectra are presented in Fig. 1.

As mentioned above, in similar experiments conducted with lower excitation wavelength (514.5 nm), no 2D band splitting was observed.[13, 14, 20-22] Huang *et al.*[19] using the 532nm excitation reported on a clear 2D splitting into two distinct components only when large strain was applied on graphene along high-symmetry directions (zigzag and armchair configurations). The intensities of both sub-peaks showed strong dependence on the incident and scattered light polarization. It is interesting to note that whereas their reported shift rates for the G$^-$ (G$^+$) modes differ by a factor of 3 (2) compared to our earlier data,[12, 13] the deviation is less pronounced for the 2D components.

An explanation for this apparent discrepancy is provided here. Figure 5(b) shows the FWHM(2D) as a function of strain for the flakes F1 and F5 excited by the 785 nm laser and for the flake F3 [12, 14] excited by the 514.5 nm (the laser polarization was parallel to the strain axis). The 2D band was fitted, for this purpose, by a single Lorentzian, which is an adequate fit for the 514.5 nm excited spectra.[13, 14] However, in the case of 785 nm excitation, a single peak fitting is not satisfactory due to pronounced splitting. Therefore, an accurate comparison of 2D peak shift rates is problematic and should be avoided. Instead, G sub-peaks should be used to assess the validity of the experimental procedure involving uniaxial tension, because their shift rates are not influenced by the excitation energy and/or by the polarization angle. The G sub-peaks shift rates of the flakes in the presented study (see also Ref. 12) are in a perfect accordance with the experimental and theoretical results presented in Ref. 13. Furthermore, the FWHM(2D) difference between the two excitation wavelengths is undisputable, despite their comparable shift rates.[12-14] Using the 514.5 nm excitation, the FWHM(2D) does not exceed even 40 cm$^{-1}$ at 1.3 % of strain, while with 785 nm excitation the FWHM(2D) reaches approx. 43 cm$^{-1}$ at strain of only 0.7%. Similar values were recorded for the flake F5 (Fig. S6, Supporting Information). It should be stressed that the dispersive behavior of the 2D peak[32], which accounts for a difference in the initial 2D band position by approx. 100 cm$^{-1}$ when excited by 514.5 and 785 nm, will alter the 2D frequency shift rate only marginally (a decrease of only 2.4 cm$^{-1}$/% for 785 nm excitation is expected[13]).

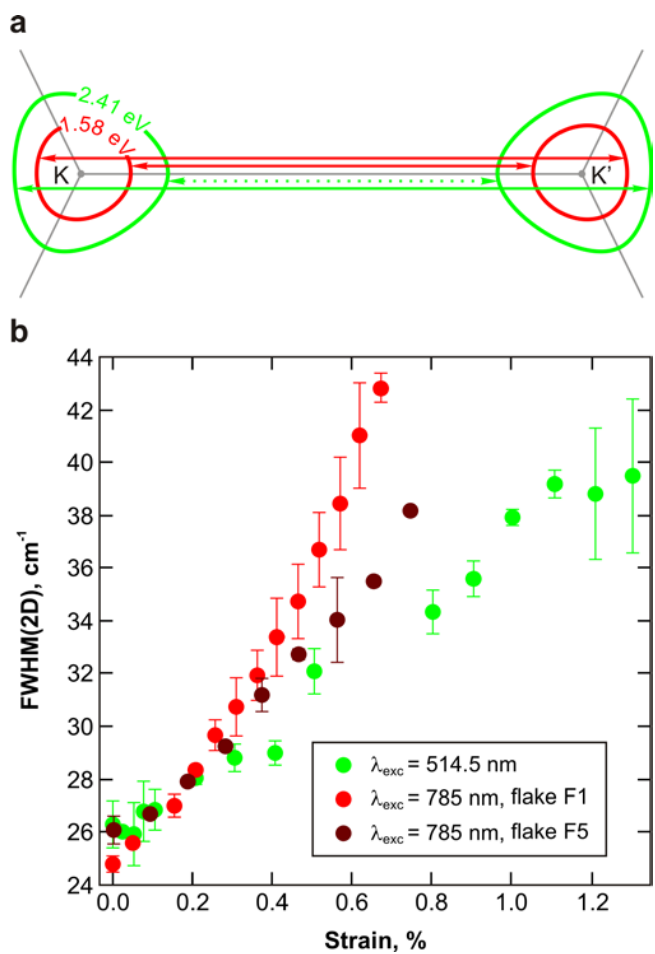

**Figure 5.** (a) Equi-excitation-energy contour plots around the K and K' points of $\pi$ electrons involved in the scattering process for 2.41 eV and 1.58 eV excitation energies. The electronic energies were calculated within the tight-binding approximation including only nearest neighbors ($\gamma_o$=-2.8 eV). In each contour the corresponding horizontal lines indicate schematically the inner and outer process. The dotted green line corresponding to the 2.41 eV contour represents a possibly weaker inner process (see text). (b) Evolution of the Raman FWHM of the 2D mode under tensile strain as measured on the graphene flakes F1 (red) and F5 (brown) (both $\lambda_{exc}$ = 785 nm) and F3 (green) ($\lambda_{exc}$ = 514.5 nm) (for details on F3 flake, see Refs. 12, 14) when fitted as a single Lorentzian. The laser polarization was in all cases parallel ($\theta_{in}$=0°) w.r.t. the strain axis.

The most striking feature of the results presented here is the qualitatively different behavior of the 2D mode for the 785nm and 514.5nm excitation. Fig. 5(a) shows electronic equi-excitation-energy contours for 1.58 eV and 2.41 eV excitation energies, derived by means of a simple nearest neighbor tight binding theory. As it is evident from Fig. 5(a) the trigonal warping effect becomes stronger and the Dirac cones more deformed as the excitation energy increases. Kürti el al.[33] calculated the Raman amplitude for $E_{laser}$=2.0 eV and 2.5eV performing a two-dimensional integration over the graphene Brillouin zone. They predicted that the highest Raman intensity occurs for the outer process, namely for transitions between parallel parts of the corresponding contours resulting to a local maximun in Raman amplitude (Fig. 5(a)). Moreover, a local maximum of much lower intensity is obtained for transitions (inner process) between parts with the highest curvature (Fig. 5(a)). However, the used phonon dispersion in Ref. 33 showed a much stronger trigonal warping than recent experiments showed,[34] which probably results in different relative intensities between inner and outer processes. Also, all calculations so far used the approximation of constant optical transition and electron-phonon matrix elements.[27, 35] Thus a theoretical interpretation of the observed intensities remains difficult. Inspecting only the electronic density of states, the $q_{in} < q_{KK'}$ phonon (inner process) involves a smaller portion of the phase space because of the anisotropy (trigonal warping) of the electron dispersion relation.[6, 7, 33] It should be mentioned that the $q \approx q_{KK'}$ phonon has a zero electron-phonon coupling for a DR transition.[36] According to Maultzsch et al.[11] the contributions by phonons having $q = q_{KK'}$ cancel out due to destructive interference. On the other hand, at lower optical transition energies such as 1.58 eV, the equi-excitation-energy contour is notably more round and the trigonal warping effect weaker (Fig. 5(a)). In that case, it seems that both inner and outer process give comparable contribution to the Raman signal. Ideally, if both electron and phonon dispersions are isotropic, the involved DR phonon wave vectors for the inner ($q_{in}$) and outer ($q_{out}$) processes should differ by the same ammount compared to the $q_{KK'}$. Since the phonon frequency depends on the absolute value of the difference $q_{out} - q_{KK'}$ or $q_{in} - q_{KK'}$, the phonon frequency for the inner and outer processes should coincide. Very recent inelastic X-ray experiments[34] showed that the equienergy contour of the TO mode in the vicinity of the K point is almost circular. Therefore, we attribute the observed complex line shape of the 2D peak for 1.58 eV excitation to the DR activation from both phonon branches along the K − M (outer process) and Γ − K (inner process) BZ directions. And it is tempting to assume that for 514.5nm excitation the inner process is weaker due to the lower joint density of states in the corresponding area of the ***k***-space and thus the fit of the 2D peak by a single Lorentzian is appropriate (Fig. 5). As already mentioned, Huang et al.[19] observed 2D splitting for uniaxial tension along only the zigzag and armchair directions, using the 532 nm excitation. They interpret their data in terms of major contribution from inner DR scattering. As evidenced above, the 2D mode excited with 785 nm (1.58 eV) is not a single peak and has a complex line-shape due to the contribution of two distinct processes (inner and outer) to the double resonance signal. The overall behaviour indicates an alteration of the inner/outer contributions to the DR scattering signal in graphene, as a function of the incident photon energy. Also, the present study will have important implications in the field of graphene based nanocomposites since measurements of the 2D band alone, for excitation energies lower than 2.41 eV (514.5 nm), may lead to errors with regards to the detailed determination state of stress or strain in the specimen.

CONCLUSIONS

In summary, we present a comprehensive experimental and theoretical analysis of the double-resonant 2D Raman mode in graphene having different orientations, under uniaxial strain. A pronounced 2D splitting into two distinct components is clearly evident in all the studied specimens using the 785 nm (1.58 eV) excitation. The results can be explained considering: (i) the strain induced asymmetry of the Brillouin zone, (ii) the additional contribution of the inner double-resonance scattering mechanism, and (iii) the incident laser polarization direction, with respect to the strain axis. Taking into account previous experiments,[13, 14] conducted with 514.5 nm (2.41 eV) excitation, our study strongly suggests that the 2D mode lineshape depends on the excitation energy. Thus, Raman measurements using various excitation wavelengths, under well defined strain conditions, are extremely important for a complete picture of the 2D mode scatering process in graphene.

METHODS

Graphene monolayers were prepared by mechanical cleavage from natural graphite (Nacional de Grafite) and transferred onto the PMMA cantilever beam covered by a ~200 nm thick layer of SU8 photoresist (SU8 2000.5, MicroChem). After placing the graphene samples, a thin layer of S1805 photoresist (Shipley) was spin-coated on the top. The top surface of the beam can be subjected to a gradient of applied strain by flexing the beam by means of an adjustable screw positioned at a distance $L$ from the fixed end. The deflection $δ$ was measured accurately using a dial gauge micrometer attached to the top surface of the beam. Furthermore, the total thickness of the beam $t$ and the flake's distance from the fixed end $x$ are taken into account for the calculation of the strain level.[14, 37] The validity of this method for measuring strains within the -1.5% to +1.5% strain range has been verified earlier.[38]

MicroRaman (InVia Reflex, Rensihaw, UK) spectra are recorded with 785 nm (1.58eV) excitation, while the laser power was kept below 0.85 mW to avoid laser induced local heating on the sample. A 100x objective with numerical aperture of 0.9 is used, and the spot size is estimated to be ~1x2 μm. The polarization of the incident light was kept either parallel ($θ_{in}=0°$) or perpendicular ($θ_{in}=90°$) to the applied strain axis, while the scattered light polarization was selected, in all cases, parallel to the strain axis ($θ_{out}=0°$). All peaks in the Raman spectra of graphene were fitted with Lorentzians. The 2D linewidths together with 2D/G relative intensities were used to identify graphene monolayers.[7, 8]

Calculations were performed with the code QUANTUM-Espresso.[39] We used a plane-wave basis set, RRKJ pseudopotentials [40] and the generalized gradient approximation in the Perdew, Burke and Ernzerhof parameterization for the exchange-correlation functional.[41] A Methfessel-Paxton broadening with a width of 0.02 Ry was used.[42] The valence electrons were expanded in a plane wave basis with an energy cutoff of 60 Ry. A 42x42x1 sampling grid was used for the integration over the Brillouin zone. The dynamical matrices were calculated on a 12x12x1 q-grid using the implemented linear-response theory. Force constants were obtained via a Fourier transformation and interpolated to obtain phonons at arbitrary points in the Brillouin zone. It should be stressed that the use of the GW approximation would improve the description of the TO phonon around the K point.[43] However the implementation is very expensive and would have to be applied on all strain directions and magnitude.

Uniaxial strain was applied via the two-dimensional strain tensor $\epsilon=((\epsilon,0),(0,-\epsilon\nu))$, where $\nu$ is the Poisson's ratio. For strain in arbitrary directions the strain tensor is rotated by $\epsilon'=R^{-1}\epsilon R$, where $R$ is the rotational matrix. After applying a finite amount of strain we relax the coordinates of the basis atoms until forces are below 0.001 Ry/a.u. and minimize the total energy with respect to the Poisson's ratio $\nu$. For small strain values we obtain a Poisson's ratio $\nu=0.164$, in excellent agreement with experimental tension measurements on pyrolytic graphite that yield a value of $\nu=0.163$.[44]

ACKNOWLEDGMENT

FORTH / ICE-HT acknowledge financial support from the Marie-Curie Transfer of Knowledge program CNTCOMP [Contract No.: MTKD-CT-2005-029876]. K.S.N. is grateful to the Royal Society and European Research Council (grant 207354 – "Graphene") for support. O.F. and L.K. further acknowledge the financial support of Czech Ministry of Education Youth and Sports (contract No. LC-510) and the Academy of Sciences of the Czech Republic (contracts IAA 400400804 and KAN 200100801). M.M. acknowledges support from EC 7th framework programme unter contract number CP-IP 228579-1. J.M. acknowledges support from the European Research Council (grant 259286).## SUPPORTING INFORMATION AVAILABLE

Details on the determination of crystallographic orientation of the studied flakes, including a Table 1. Optical micrographs of the studied flakes (Figure S1), a sketch of the graphene crystal lattice (Figure S2), calculated 2D frequencies for the individual outer and inner processes along different K − K' paths, for $\varphi = 10°$ (Figure S3), 2D band Raman spectra of graphene flake F4 (Figure S4), simulated 2D Raman spectra for $\varphi = 10°$ (Figure S5), 2D band Raman spectra of graphene flake F5 (Figure S6), splitting of the Raman 2D band under tensile strain as measured on the graphene flake F5 (Figure S7), simulated Raman 2D band under uniaxial tension for $\varphi = 20°$ showing all six calculated components (Figure S8). This material is available free of charge via the Internet at http://pubs.acs.org.

## FIGURE CAPTIONS

**Figure 1.** 2D mode Raman spectra of graphene flake excited at 785 nm under uniaxial tension. The spectra in the left and right panels were acquired with laser polarization parallel ($\theta_{in}=0°$) and perpendicular ($\theta_{in}=90°$) respectively, w.r.t. the strain axis. Data were recorded around the center of the flake F1. The original measurements are plotted as points. The solid curves are the best Lorentzian fits to the experimental spectra while the dashed lines indicate the two components (see text). The tension level is labeled at each curve.

**Figure 2.** (a) The first Brillouin zone of uniaxially strained graphene for {$\epsilon=0.3$, $\varphi = 20°$}. The $\boldsymbol{b_1}$ and $\boldsymbol{b_2}$ denote the primitive vectors of the reciprocal lattice. The high symmetry points and the irreducible parts (shaded areas) of the Brillouin zone are depicted. (b) Incoming double resonance mechanisms (left: outer process and right: inner process) plotted into the band structure scheme of graphene along the high symmetry KMK' path. For simplicity we omit the equivalent process for hole-phonon scattering. (c) Calculated 2D frequencies for the outer (o) and inner (i) processes along different K – K' paths (Fig. 2(a)), for laser excitation 785nm and $\varphi = 20°$. The blue solid (green open) symbols correspond to the outer (inner) process. The lines are guides to the eye.

**Figure 3.** Simulated Raman spectra in the 2D frequency region calculated for 785 nm excitation of a graphene flake with crystal lattice orientation $\varphi = 20°$ and for various tensile strain levels. The spectra in the left and right panels were calculated for laser polarization parallel ($\theta_{in}=0°$) and perpendicular ($\theta_{in}=90°$) w.r.t. the strain axis, respectively. Dashed lines indicate fits with two Lorentzian components to the calculated spectra.

**Figure 4.** Splitting of the Raman 2D mode under tensile strain as measured on the graphene flake F1. Red and black colors indicate laser polarization parallel ($\theta_{in}=0°$) and perpendicular ($\theta_{in}=90°$) w.r.t. the strain axis, respectively. Full and empty squares of the same color indicate the frequency of the two main Lorentzian components of the 2D mode. Solid lines (only for $\theta_{in}=0°$) represent linear fits where all measurements on the flake have been taken into account. The data points for $\theta_{in}=90°$ were offset by -4 cm$^{-1}$ in order to provide a better visual comparison to data acquired with $\theta_{in}=0°$. The original spectra are presented in Fig. 1.

**Figure 5.** (a) Equi-excitation-energy contour plots around the K and K' points of $\pi$ electrons involved in the scattering process for 2.41 eV and 1.58 eV excitation energies. The electronic energies were calculated within the tight-binding approximation including only nearest neighbors ($\gamma_o$=-2.8 eV). In each contour the corresponding horizontal lines indicate schematically the inner and outer process. The dotted green line corresponding to the 2.41 eV contour represents a possibly weaker inner process (see text). (b) Evolution of the Raman FWHM of the 2D mode under tensile strain as measured on the graphene flakes F1 (red) and F5 (brown) (both $\lambda_{exc}$ = 785 nm) and F3 (green) ($\lambda_{exc}$ = 514.5 nm) (for details on F3 flake, see Refs. 12, 14) when fitted as a single Lorentzian. The laser polarization was in all cases parallel ($\theta_{in}$=0°) w.r.t. the strain axis.

26
(6)     Ferrari, A. C. Raman Spectroscopy of Graphene and Graphite: Disorder, Electron-Phonon Coupling, Doping and Nonadiabatic Effects. *Solid State Commun.* **2007,** *143*, 47-57.

(7)     Ferrari, A. C.; Meyer, J. C.; Scardaci, V.; Casiraghi, C.; Lazzeri, M.; Mauri, F.; Piscanec, S.; Jiang, D.; Novoselov, K. S.; Roth, S., et al. Raman Spectrum of Graphene and Graphene Layers. *Phys. Rev. Lett.* **2006,** *97*, 187401.

(8)     Malard, L. M.; Pimenta, M. A.; Dresselhaus, G.; Dresselhaus, M. S. Raman Spectroscopy in Graphene. *Phys. Rep.* **2009,** *473*, 51-87.

(9)     Ni, Z. H.; Wang, H. M.; Kasim, J.; Fan, H. M.; Yu, T.; Wu, Y. H.; Feng, Y. P.; Shen, Z. X. Graphene Thickness Determination Using Reflection and Contrast Spectroscopy. *Nano Lett.* **2007,** *7*, 2758-2763.

(10)    Tuinstra, F.; Koenig, J. L. Raman Spectrum of Graphite. *J. Chem. Phys.* **1970,** *53*, 1126-1130.

(11)    Maultzsch, J.; Reich, S.; Thomsen, C. Double-Resonant Raman Scattering in Graphite: Interference Effects, Selection Rules, and Phonon Dispersion. *Phys. Rev. B* **2004,** *70*, 155403.

(12)    Frank, O.; Tsoukleri, G.; Parthenios, J.; Papagelis, K.; Riaz, I.; Jalil, R.; Novoselov, K. S.; Galiotis, C. Compression Behavior of Single-Layer Graphenes. *Acs Nano* **2010,** *4*, 3131-3138.

(13)    Mohiuddin, T. M. G.; Lombardo, A.; Nair, R. R.; Bonetti, A.; Savini, G.; Jalil, R.; Bonini, N.; Basko, D. M.; Galiotis, C.; Marzari, N., et al. Uniaxial Strain in Graphene by Raman Spectroscopy: G Peak Splitting, Grueneisen Parameters, and Sample Orientation. *Phys. Rev. B* **2009,** *79*, 205433-8.

(14)    Tsoukleri, G.; Parthenios, J.; Papagelis, K.; Jalil, R.; Ferrari, A. C.; Geim, A. K.; Novoselov, K. S.; Galiotis, C. Subjecting a Graphene Monolayer to Tension and Compression. *Small* **2009,** *5*, 2397-2402.

SYNOPSIS_TOC

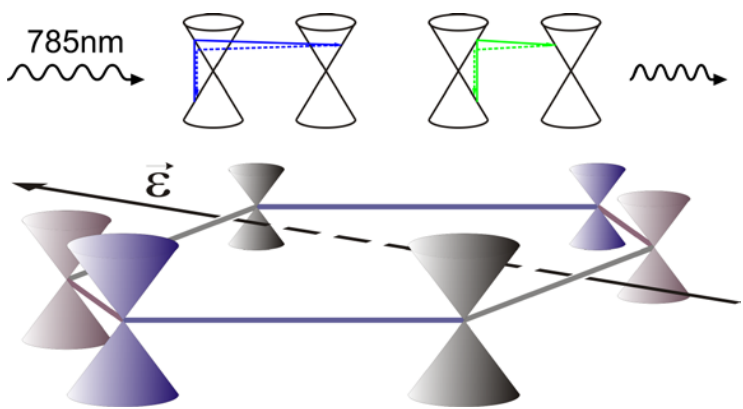